\documentclass[preprint]{aastex}
\shorttitle{Starburst Close to the Nucleus of NGC\,1097 }
\shortauthors{Storchi-Bergmann et al.}

\begin{document}


\title{Evidence for a Starburst within 9\,pc of the Nucleus of NGC\,1097\altaffilmark{1}}


\author{T. Storchi-Bergmann\altaffilmark{2}, R. S. Nemmen\altaffilmark{2}, P. F. Spinelli\altaffilmark{2}, M. Eracleous\altaffilmark{3},\\ A. S. Wilson\altaffilmark{4}, A. V. 
Filippenko\altaffilmark{5} and M. Livio\altaffilmark{6}} 
%\affil{Instituto de F\'{\i}sica, UFRGS, Porto Alegre, RS, Brazil}
\email{thaisa@if.ufrgs.br}

%\author{M. Eracleous}
%\affil{Department of Astronomy and Astrophysics, Pennsylvania State University,
%525 Davey Laboratory, University Park, PA 16802}
%\email{mce@astro.psu.edu}

%\author{A. S. Wilson\altaffilmark{3}}
%\affil{Department of Astronomy, University of Maryland, College Park, MD 20742}
%\email{wilson@astro.umd.edu}

%\author{Alexei V. Filippenko\altaffilmark{4}}
%\affil{Department of Astronomy, University of California,
%601 Campbell Hall, Berkeley, CA 94720-3411}
%\email{alex@astro.berkeley.edu}

%\and

%\author{M. Livio\altaffilmark{5}}
%\affil{Space Telescope Science Institute, Baltimore, MD 21218}
%\email{mlivio@stsci.edu}

\altaffiltext{1}{Based on observations with the NASA/ESA {\it Hubble Space Telescope} at the
Space Telescope Science Institute, which is operated by the Association of Universities
for Research in Astronomy, Inc., under NASA contract NAS5-26555.}
\altaffiltext{2}{Instituto de F\'{\i}sica, UFRGS, C.P. 15051, Porto Alegre, RS, Brazil}
\altaffiltext{3}{Department of Astronomy and Astrophysics, Pennsylvania State University,
PA 16802; mce@astro.psu.edu}
\altaffiltext{4}{Department of Astronomy, University of Maryland, College Park, MD 20742;
wilson@astro.umd.edu}
\altaffiltext{5}{Department of Astronomy, University of California,
Berkeley, CA 94720-3411; alex@astro.berkeley.edu }
\altaffiltext{6}{Space Telescope Science Institute, Baltimore, MD 21218; mlivio@stsci.edu}



\begin{abstract}
%In 1993, we reported the unexpected discovery of a 
%broad (FWHM $\approx 10,000$ km s$^{-1}$),
%double-peaked H$\alpha$ emission line in the nucleus
%of NGC\,1097, well-reproduced by an
%accretion-disk model. Follow-up observations mapped variations
%in the double-peaked line which provided evidence
%against other possible origins
%and constrained the accretion-disk model.

We report evidence for a recent burst of star formation located 
within 9\,pc of the active nucleus of NGC\,1097. The observational
signatures of the starburst include UV absorption lines
and continuum emission from young stars observed in a
small-aperture {\it HST} spectrum. The burst
is $\le$ a few $\times$ 10$^6$\,yr old, has a mass of 
$\sim\,10^6$\,M$_\odot$, an observed luminosity of 1.5 $\times$ 10$^7$
L$_\odot$ and is obscured by $A_V\approx 3$ mag.
The importance of this finding is two-fold:  (1) the proximity of the
starburst to the active nucleus and thus possible association
with it; (2) its obscuration by and apparent
association with a dusty absorbing medium, while 
%the nuclear continuum and 
the broad emission lines appear unobscured,
suggesting that the starburst could be embedded in a circumnuclear
torus as predicted in the Unified Model of active galactic nuclei.
 
\end{abstract}


%% subject header.  Objects should be in the appropriate "individual"
%% headers (e.g. quasars: individual, stars: individual, etc.) with the
%% additional provision that the total number of headers, including each
%% individual object, not exceed six.  The \objectname{} macro, and its
%% alias \object{}, is used to mark each object.  The macro takes the object
%% name as its primary argument.  This name will appear in the paper
%% and serve as the link's anchor in the electronic edition if the name
%% is recognized by the data centers.  The macro also takes an optional
%% argument in parentheses in cases where the data center identification
%% differs from what is to be printed in the paper.

\keywords{galaxies: active ---
galaxies: individual (NGC 1097) ---
accretion disk --- starburst}

%% Ex.: It has been predicted by theory \citep{hen61,lyn68,spi85}, but

\section{Introduction}

The standard model of an active galactic nucleus (AGN) consists
of a supermassive black hole being fed via an accretion disk.
%possibly surrounded by dense clouds --- the broad-line region.
The model also postulates the presence of a dusty molecular torus
surrounding this inner region, and blocking it from view in AGNs
where the line-of-sight intercepts the torus \citep{ant93}.

For several years, the only plausible signature of the accretion disk was the
``big blue bump,'' observed in the ultraviolet (UV) spectra of active galaxies,
which was supposed to be the thermal emission from the disk 
\citep{shi78,mal82,kor99}.
A more telling feature, such as the kinematic
signature of rotating gas, in the form of double-peaked
line emission as observed  in cataclysmic variables,
was only discovered  in optical spectra in the 1980s
\citep{hf88,chf89,hal90}, and systematic searches
were later published by \citet{mce94,eh03} and by \citet{str03}.
Broad, double-peaked emission lines from the nucleus of NGC\,1097
were reported  by \citet{sto93}, and similar lines were found in other 
nearby galaxies with the
{\it Hubble Space Telescope (HST)}
\citep{bow96,shi00,ho00,bar01}.
 
NGC\,1097 is particularly interesting 
because the double-peaked line originates in a
low-luminosity AGN and is a transient phenomenon, as it was not
seen in previous observations.
These characteristics indicate that we would have
a good chance of observing temporal variations in a reasonable
(not too long) time interval. Indeed, monitoring of the double-peaked
emission over the years 1992--2001 \citep{sto97,sto03} has put further 
constraints on the accretion-disk model and provided evidence against 
other possible origins for the double-peaked emission in this
particular object, such as
double jets and a binary black hole \citep{sto95,liv96,mce97}.

In this {\it Letter}, we report the presence of absorption features
in small-aperture  $HST$ UV spectra of NGC\,1097, and argue that they are
characteristic of young O and B stars and of a low-ionization interstellar medium,
all within a radius of at most 9\,pc from the nucleus. Such dimensions are
typical of those of star clusters and also of those expected
for obscuring tori in the standard model of AGNs. Besides the
absorption features, the UV continuum also reveals
the spectral energy distribution of a
young starburst extinguished by a dusty medium.

\section{Observations}

%% In a manner similar to \objectname authors can provide links to dataset
%% hosted at participating data centers via the \dataset{} command.  The
%% second curly bracket argument is printed in the text while the first
%% parentheses argument serves as the valid data set identifier.  Large
%% lists of data set are best provided in a table (see Table 3 for an example).
%% Valid data set identifiers should be obtained from the data center that
%% is currently hosting the data. 

%% In this section, we use  the \subsection command to set off
%% a subsection.  \footnote is used to insert a footnote to the text.


%% This section also includes several of the displayed math environments
%% mentioned in the Author Guide.

UV and optical spectra of  NGC\,1097 were obtained with the Space 
Telescope Imaging Spectrograph (STIS) aboard $HST$ on 2001 February 7 
and 11 UT, using a $0\farcs2 \times 52^{\prime\prime}$ slit 
and the  G140L, G230L, G430L,
and G750L gratings. These gratings have resolving powers $500\le R\le 1000$,
which allow the separation of absorption features in the galaxy from
those in the Milky Way. The exposure times were 8302\,s for each of the
two UV gratings, 1008\,s for grating G430L, and 936\,s for grating G750L.
In order to obtain the best possible angular resolution achievable with the
observations, we extracted a nuclear spectrum within a
window of $0\farcs2$, corresponding to 
a 9\,pc radius region centered on the nucleus.
We then combined the various spectral segments to construct
one UV/optical spectrum spanning the spectral range 1100--9000~\AA.

\begin{figure}
%\epsscale{.80}
\plotone{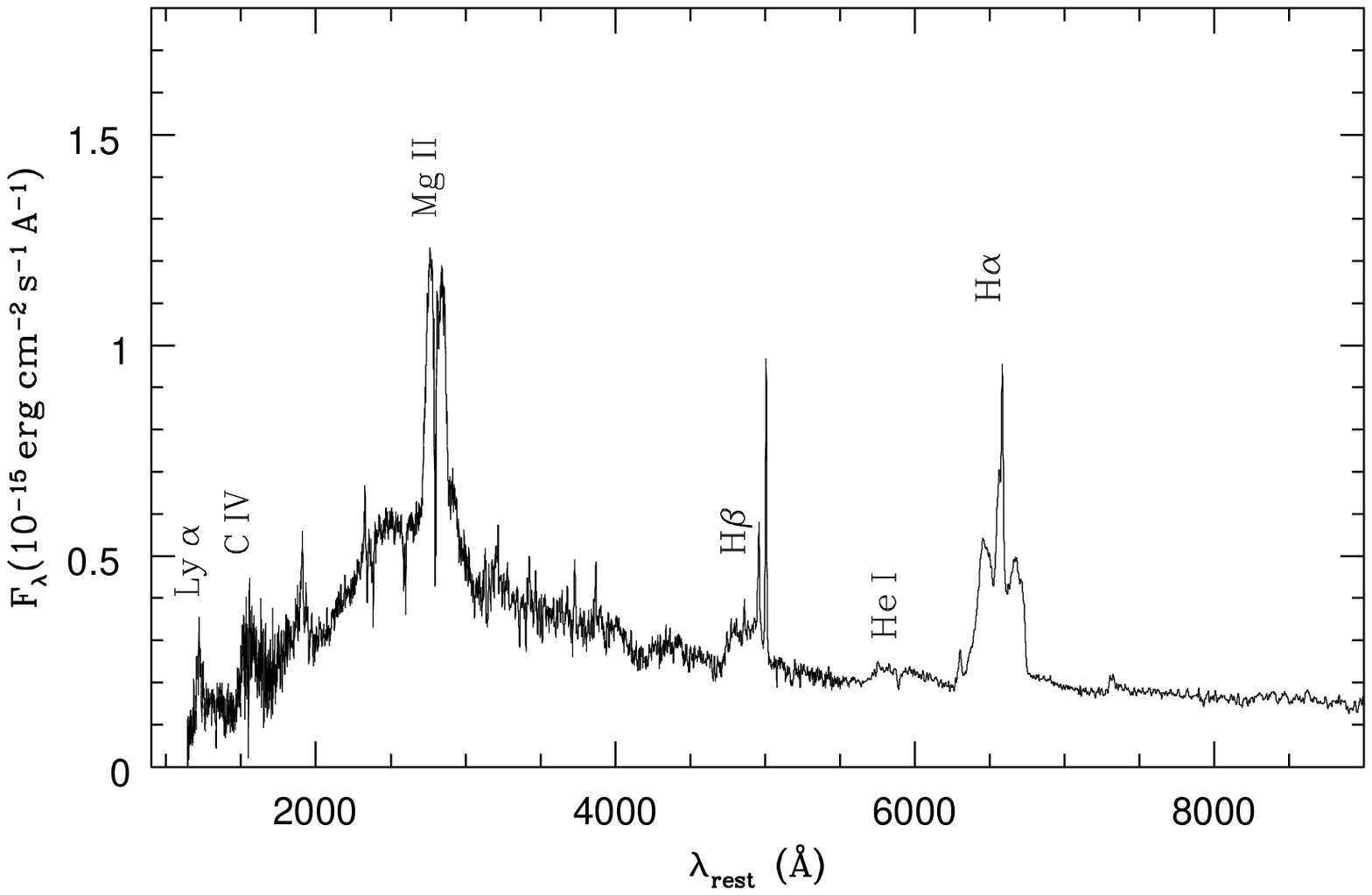}
\caption{The UV + optical spectrum of NGC\,1097 from a region within
$0\farcs 1$ (9\,pc at the galaxy) of the nucleus, in the rest
frame of the galaxy. The double-peaked
and other broad lines are identified.\label{fig1}}
\end{figure}

The combined spectrum was corrected for the foreground reddening
$A_V = 0.088$ mag from \citet{sch98}. The spectrum, in the
rest frame of the galaxy, is shown in Figure \ref{fig1},
where we have edited out the geocoronal Ly\,$\alpha$
and O\,I\,$\lambda$1302.08 emission lines. The main features readily
observed in this spectrum are a ``hump''  in the UV continuum peaking
near 2500~\AA, and several double-peaked emission lines, identified
in the figure: Mg\,II\,$\lambda$2800, H$\beta$,
He\,I\,$\lambda$5876, and H\,$\alpha$. Broad components are also observed
in Ly\,$\alpha$, C\,IV\,$\lambda$1550, and C\,III]$\lambda$1909, although the 
double-peaked nature is not clear because of the lower 
signal-to-noise ratio.

\section{A Close Look at the UV Spectrum: Absorption Features}

The UV spectra of nuclei with broad double-peaked emission
lines (hereafter dubbed ``doublepeakers'') show a number of
common absorption features in the spectral range 2000--3000~\AA,
mostly metastable Fe\,II absorption lines \citep{mce02,mce04}.
The absorption spectrum of NGC\,1097 in the above spectral
range is similar to that of the prototypical
doublepeaker, Arp\,102B \citep{hal96}, which \citet{mce03} have
proposed to originate in thin sheets or filaments embedded in an
outflowing wind that overlays the accretion disk (and thus is close to it,
within roughly 5000 gravitational radii). \citet{mce04}
propose that many doublepeakers may have these filaments, and NGC\,1097
probably also fits within this scenario.
%, as discussed at the end of this section.

\begin{figure}
%\epsscale{.80}
\plotone{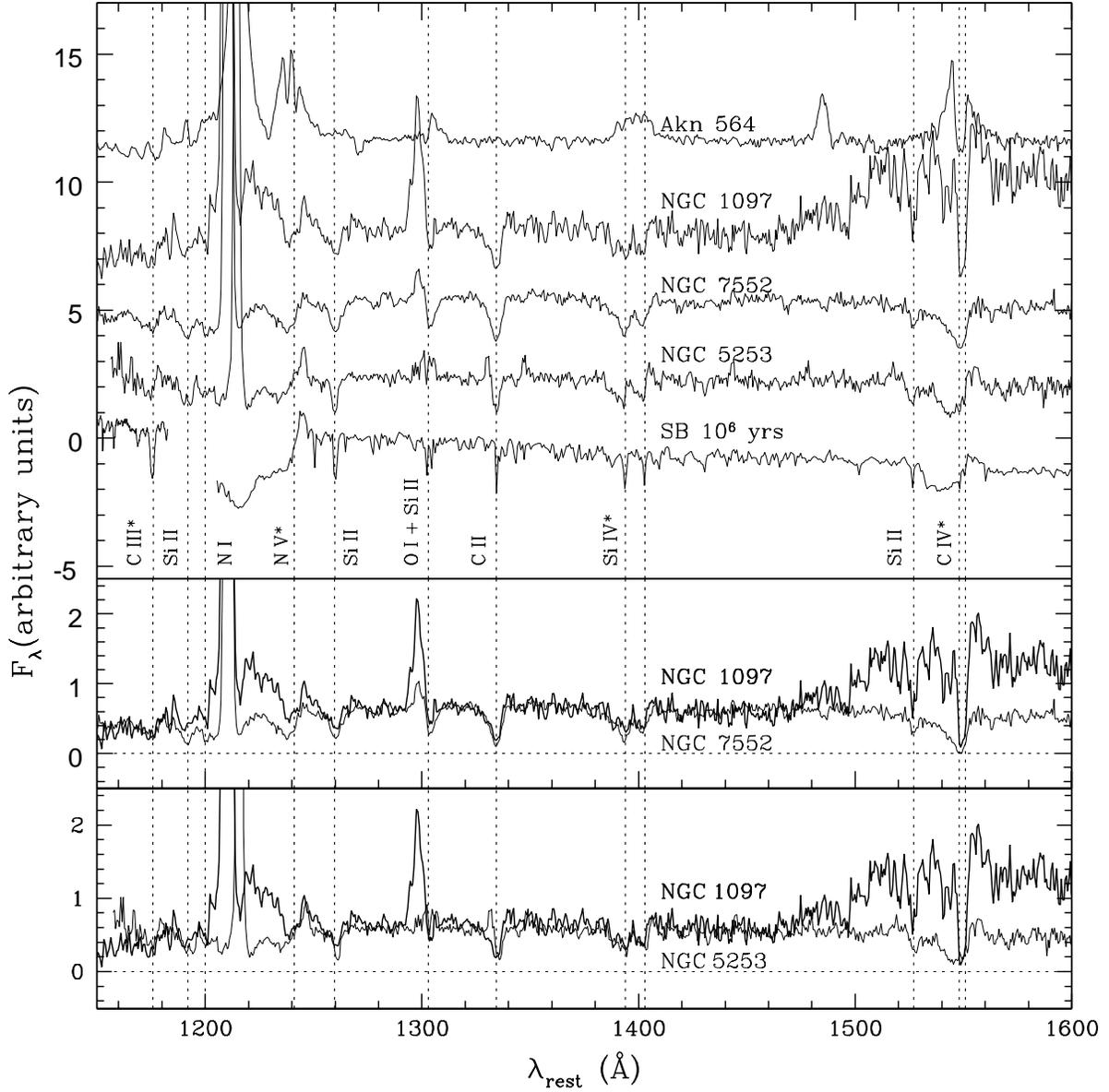}
\caption{{\it Top:} The UV spectrum of NGC\,1097 compared to those
of the Seyfert\,1 galaxy Akn\,564 and  the starburst nuclei of
NGC\,7552 and NGC\,5253. All spectra are in the rest frame of the galaxies.
Also included is a synthetic spectrum of a 10$^6\,M_\odot$,
10$^6$\,yr-old starburst obtained using the code Starburst\,99 \citep{lei99}.
The absorption lines are identified at the bottom. An asterisk means that
the line is present in the atmosphere of young stars. {\it Middle:}
The NGC\,1097 spectrum superimposed on the scaled NGC\,7552 
spectrum. {\it Bottom:}
The NGC\,1097 spectrum superimposed on the scaled NGC\,5253
spectrum.\label{fig2}}
\end{figure}

Here we will focus on the range 1100--1600~\AA\
of the nuclear spectrum  of NGC\,1097. 
%Unlike the other doublepeakers,
In this spectral range, 
NGC\,1097 shows a number of absorption lines, some of which are superimposed
on the broad Ly\,$\alpha$ and C\,IV emission lines, as illustrated in
the top panel of Figure \ref{fig2}. The absorption features are identified at
the bottom of the figure.
We include in this figure, for comparison, the STIS spectrum
of the Seyfert galaxy Akn\,564 (representative of Seyfert galaxy spectra)
and of the starburst nuclei of NGC\,7552 and NGC\,5253, 
all in the rest frame of the galaxies.
Upon inspection of the UV spectra of many other
Seyferts and starbursts, we conclude that the absorption
features in the UV spectrum of NGC\,1097
are not common in AGNs, but are frequently
observed in the spectra of starburst galaxies.

We also include in Figure \ref{fig2}
a synthetic spectrum of a starburst 
obtained using the code Starburst\,99 \citep{lei99},
with the following parameters:
instantaneous burst of mass 10$^6~M_\odot$, age  10$^6$\,yr,
Salpeter initial mass function (IMF), lower and upper limits for
the IMF of $1\,M_\odot$ and $100\,M_\odot$ (respectively), and  solar metallicity.
The spectra of Figure \ref{fig2} have been normalized at 1350~\AA\
to allow a better comparison among them, and have been shifted vertically
for clarity.

%*** Thaisa, below I fixed what I think is an error. You had said
% ``For example, as the exposure time of the observation of NGC 1097
% is longer, the geocoronal emissions are stronger relative to the
% galaxy spectrum...''   Well, both the geocoronal emission and the
% galaxy light would grow in proportion to the exposure time. So I
% think what you really meant is that the geocoronal line is simply
% bright relative to the stellar continuum. I have altered the wording
% accordingly.

In order to further  emphasize the similarity of the nuclear UV
spectrum of NGC\,1097 with those of starbursts, we show in the lower
panels of Figure \ref{fig2} the NGC\,1097 spectra on top of 
those of NGC\,7552 and NGC\,5253. We decided to keep in the spectra  
the geocoronal Ly\,$\alpha$ and
O~I $\lambda$1302 emission lines, in order to avoid introducing
artificial discrepancies. 
%(For example, the geocoronal emission in 
%the NGC\,1097 spectrum is stronger relative to the galaxy spectrum
%than in the cases of NGC\,7552 and NGC\,5253; in particular, 
%the O~I emission partially fills the galaxy absorption),
%It can be seen from this
%figure that all the absorption lines in the spectra of NGC\,7552
%and NGC\,5253 are also present in the spectrum of NGC\,1097, 
%the only difference
%being the absence of the broad Ly\,$\alpha$ and
%C\,IV emission-line components in NGC\,7552 and NGC\,5253.
%It is also clear that
%the absorption features in the spectrum of NGC\,1097 are not as
%deep as those in NGC\,7552 because, as discussed in the next section,
%there is an additional featureless continuum contributing
%to the UV light in NGC\,1097.

The following stellar absorption features, characteristic
of the spectra of early-type stars (O, B), are observed
in the spectra of NGC\,1097, NGC\,7552, and NGC\,5253 \citep{kin93,vaz04}:
C\,III\,$\lambda$1175.7, N\,V\,$\lambda\lambda$1238.8,\,1242.8, 
Si\,IV\,$\lambda\lambda$1393.8,\,1402.8, and
C\,IV\,$\lambda\lambda$1548.2,\,1550.8. The last three
are resonance lines and can also originate in the
interstellar medium. Other interstellar lines present
in the spectra are Si\,II\,$\lambda\lambda$1190.4,
1193.3, N\,I\,$\lambda$1199.9, Si\,II\,$\lambda\lambda$1259.5,\,1260.4, 
O\,I\,$\lambda$1302.1, and Si\,II\,$\lambda$1304.4 
(the last two partially filled by the geocoronal O~I emission line), 
C\,II\,$\lambda\lambda$1334.5,\,1335.7, and Si\,II\,$\lambda$1526.7. 
There are also in the nuclear spectrum of
NGC\,1097 a few absorption lines from our Galaxy, but which can be
separated from the absorption lines of NGC\,1097 because of the redshift
of the latter. 
%Particularly conspicuous is the C\,IV absorption from our 
%Galaxy, appearing as a doublet to the blue side of the NGC\,1097 C\,IV 
%absorption (Fig. \ref{fig2}). 
%In the case of NGC\,7552, the Milky Way
%absorption seems to be blended with that of the galaxy, contributing
%to its strong blue wing.

Among the stellar absorption features, only 
C\,III\,$\lambda$1175 is not a resonance line and 
must originate in the atmosphere of young stars.
This line is sharp and deep in the starburst synthetic
spectrum, but is broader and shallower in the
spectra of the actual starbursts NGC\,5253 and NGC\,7552.
In the NGC\,1097 spectrum, it has a similar profile 
to those observed in the starburst galaxies,
as can be seen in Fig. \ref{fig2}.
The presence of young stars is also supported
by the P-Cygni profile of the line N\,V\,$\lambda$1240,
which can be clearly observed in NGC\,1097,
NGC\,5253 and in the synthetic starburst spectra, although only
a hint of it appears in the NGC\,7552 spectrum. Notice, in particular,
the similarity between the emission portion of this line in
NGC\,1097 and NGC\,5253 (Fig \ref{fig2}).

A P-Cygni profile is also observed in the 
C\,IV\,$\lambda$1550 line of the NGC\,5253 and synthetic
starburst spectra, but not in NGC\,7552. In NGC\,1097, there
is only a hint of a P-Cygni  profile in the C\,IV line,
superimposed on the broad emission-line profile.
The large depth of the absorption, which reaches
practically zero flux, indicates that this line probably
originates not exclusively in the stellar atmospheres 
of young stars. A large depth is also observed in the 
Mg\,II\,$\lambda$2800 absorption.
These two deep absorption lines
suggest that some of the absorbing gas
is along our line of sight to the AGN. We thus conclude that
at least part of the absorption must come from a region very close to the
AGN. The absorber could be the starburst superwind whose filaments cross our
line of sight to the nucleus [e.g. \citet{heck00}], or 
filaments embedded in an accretion-disk wind [as is the 
case favored for the other double peakers \citep{mce04}], or both.

%We conclude that both stellar absorption features
%characteristic of young O and B
%stars and interstellar absorption features frequently observed
%in the spectra of starbursts are clearly present in the UV spectrum
%of the inner 9\,pc of NGC\,1097, suggesting that there
%is a starburst close to the nucleus of NGC\,1097.

%*** Thaisa, I don't understand the reasoning given below. Why
% can't the starburst alone be responsible for all of the absorption?
% As long as the absorbing region is outside the BLR, it is possible
% to get zero flux in the line center, if the absorbing region has
% sufficiently high optical depth. Please alter, or clarify.

\section{The UV/Optical Continuum and the Role of Reddening}

Is the spectral energy distribution (SED) 
compatible with the presence of a young starburst in the
nucleus of NGC\,1097? In order to answer this question, 
we have used the code Starburst\,99 \citep{lei99} to construct model
SEDs of continuous and single bursts of star formation
of different ages. Inspection of the nuclear continuum of NGC\,1097 in
Figure \ref{fig1} shows that the flux increases from the optical to the UV,
as expected for a young starburst, but then, at $\sim$2500\,\AA,
decreases to shorter wavelengths. The only way to have this flux
decrease in the UV is through reddening.

We have tried different combinations of starburst models and reddening laws.
For the starbursts, we have also experimented with different slopes for the
initial mass function (IMF), and our conclusions are as follows.
Continuous bursts provide too much flux in the optical, as do
single bursts older than a few times 10$^6$~yr. When combined with any of the known
reddening laws, they fail to produce the observed shape of the UV
SED. The most suitable starburst model is the one of 
a single burst of age 10$^6$\,yr.
The SEDs of younger bursts or bursts with different IMFs are only
slightly different from that of a single burst of age 10$^6$~yr, and do
not provide significant improvement to the fit of the observed spectrum.
We have also concluded that models including the contribution
of the gas produce a strong Balmer jump in
emission which is not observed in the data. 
The best fit to the NGC\,1097 continuum we could
obtain from the available synthetic spectra
includes only the stellar contribution to the continuum.

In order to produce the observed flux decrease in the UV, we have
tried the Calzetti reddening law \citep{cal94},
the Milky Way (MW) reddening law \citep{ccm89},
the Large Magellanic Cloud (LMC) reddening law \citep{kor81,pei92}, and
the Small Magellanic Cloud (SMC) reddening law \citep{bou85,pei92}.
The Calzetti reddening law did not have
enough UV extinction to reproduce the data,  both the MW and
LMC laws introduced the characteristic 2200~\AA\ absorption
feature which is not present in the data, while the SMC
reddening law produced better agreement with the data
for $A_V = 3$ mag and $R = 3.1$. Although it is surprising that
the best reddening law is that of the SMC (as NGC\,1097 is a much
more luminous and a more metal-rich galaxy than the SMC), we point out that
extinction by an ``SMC-like'' reddening law has also been
found to best explain the SEDs of %273 of the 
red quasars in the SDSS quasar survey \citep{ric03}.

\begin{figure}
%\epsscale{.80}
\plotone{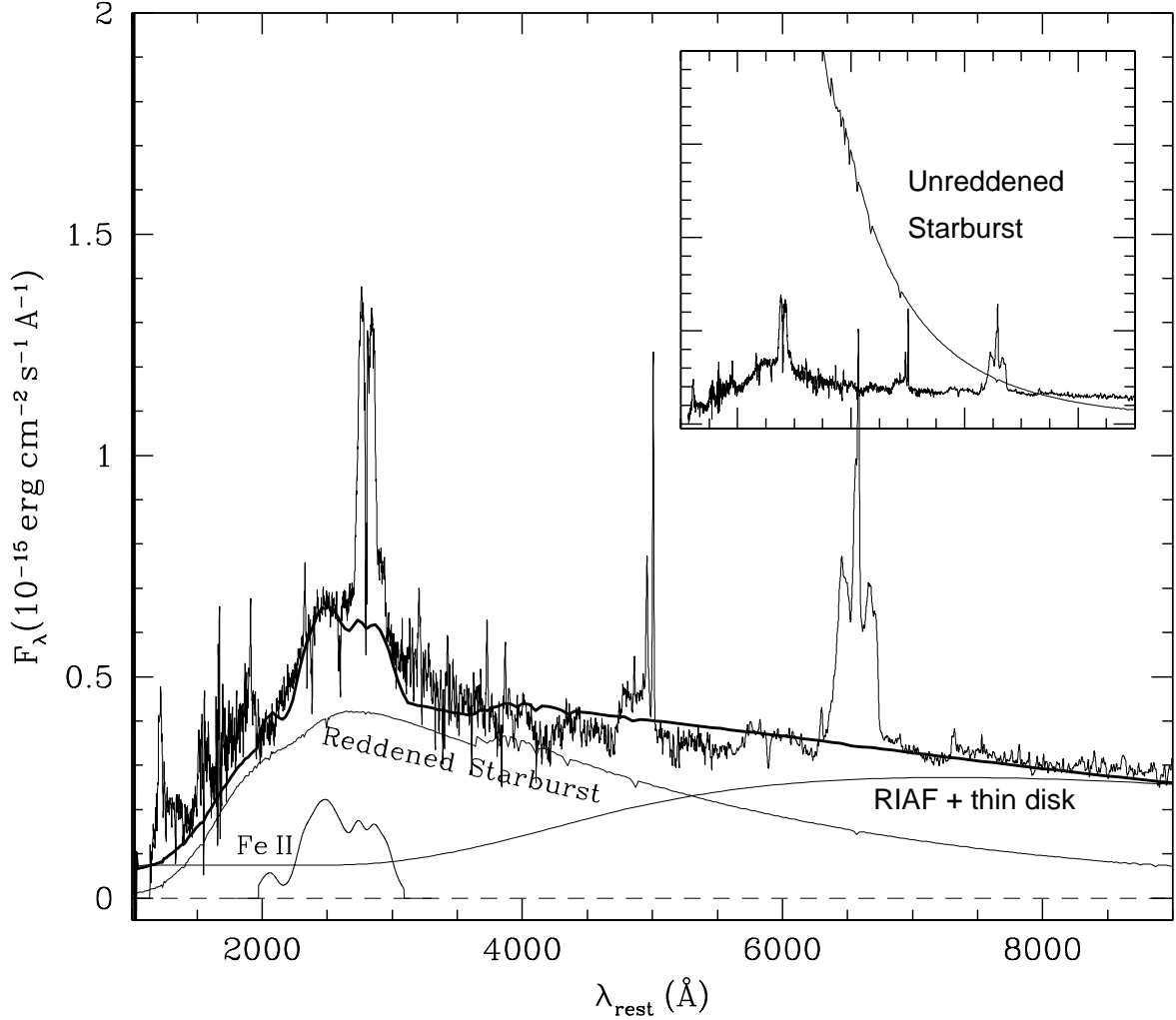}
\caption{Thin lines show the nuclear spectrum of NGC\,1097, together with the
synthetic spectrum of the 10$^6$\,yr-old starburst reddened by $A_V = 3$ mag,
the RIAF + thin-disk model, and the Fe\,II model. The thick line is
the model continuum obtained combining these three components.
The inset shows the unreddened spectrum of the starburst
compared to the NGC\,1097 spectrum.\label{fig3}}
\end{figure}

Not only the SED but also the flux of the reddened 10$^6$\,yr
starburst is in approximate agreement with the observations, indicating
that the mass of the starburst is roughly the one
adopted in the model, 10$^6\,M_\odot$. This is illustrated
in Figure \ref{fig3}, where we show the NGC\,1097
spectrum compared to the reddened starburst SED.
An inset in the figure shows the SED of the unreddened starburst
in comparison with the galaxy spectrum in order to illustrate
the effect of reddening. Figure \ref{fig3} also shows the contribution of the
continuum emitted by the AGN, borrowed from a
paper in preparation \citep{nem05}, but with preliminary results already
published \citep{nem04}. In \citet{nem05} the nuclear SED
-- from X-ray to radio wavelengths -- is found to be well reproduced by a model
consisting of the emission from a radiatively-inefficient accretion
flow (RIAF) plus that of a thin accretion disk. Although this
model produces a good overall description of the SED, it does
not reproduce the excess in the UV, as illustrated in Figure \ref{fig3}.
We point out that in \citet{nem05}, the thin-disk emission is
self-consistently derived from the incidence of the RIAF
radiation on the disk. The thin-disk continuum emission peaks
in the optical/infrared, and cannot reproduce the excess UV emission.

%The NGC\,1097 spectrum shown in
%Figure \ref{fig3} was extracted using  a somewhat larger aperture 
%(0.6$^{\prime\prime}$)
%than the 0.2$^{\prime\prime}$ of the previous spectrum,
%as we verified that the broad lines could still be detected
%outside the 0.2$^{\prime\prime}$ aperture, while none was detected
%beyond 0.3$^{\prime\prime}$ from the nucleus, and we wanted to make
%sure we were including the complete nuclear flux.
%The difference between this latter spectrum and the
%one in Figure 1 is nevertheless minimal.

\subsection{Contribution of UV Fe~II Emission}

Although the introduction of a 
reddened starburst continuum can approximately 
reproduce the UV  and its decrease to shorter wavelengths,
there is still a small remaining ``bump'' around 2500\,\AA\ which is not
reproduced by the RIAF + thin disk + reddened starburst (Fig. \ref{fig3}).
In this wavelength region there is a blend of Fe~II
emission lines which is frequently present in the UV spectra of
Seyfert~1 galaxies and quasars \citep{wills85,ves01}.

In order to investigate the possibility that the above bump
is due to the blend of Fe lines, we have used an archival
spectrum of I~Zw~1 (kindly provided to us by Karen Leighly)
as a template for the Fe emission. Although this template covers only
the 2000--3000\,\AA\ spectral region, it includes
the most intense Fe\,II emission lines. [The 
contribution of Fe\,II lines in the optical is much smaller
-- with peak intensities $\sim$10\% of that in the
UV \citep{bal04,sig03}. Detailed modelling of the optical
Fe\,II emission is beyond the scope of this work.]

Under the assumption that the Fe\,II
emission is coming from the accretion disk, we have convolved the
template with the double-peaked profile of the emission-lines.
Combining this component with the others, we could successfully
reproduce the remaining bump around 2500\,\AA.
This is also illustrated in Fig. \ref{fig3}, where
we further include the Fe\,II template and a model (thick line)
for the NGC\,1097 continuum comprising all the 
components: the reddened starburst, the RIAF + thin disk model
and the Fe\,II template, properly scaled to fit the data.
The reddened starburst model was scaled by 0.85,
corresponding to a UV + optical
flux of 1.7$\times$10$^{-12}$\,ergs\,cm$^{-2}$\,s$^{-1}$ and
luminosity of 1.5$\times$10$^7$\,L$_\odot$ for an
adopted distance to NGC\,1097 of 17\,Mpc.

Finally, we point out that it is not our goal to produce a detailed
fit to the spectrum, as there are many uncertainties.
For example, the reddening law may be somewhat different
from that of the SMC, and the starburst properties may also differ
from those of the Starburst\,99 model.

\section{Concluding Remarks}

We have shown evidence for an obscured starburst located within 9~pc
of the active nucleus of the galaxy NGC\,1097.
The importance of this finding is two-fold: (1) the proximity
of a young starburst to the nucleus and thus possible association
with it; and (2) the fact that the
starburst is obscured, but the AGN is not.

The maximum distance between the starburst and the nucleus
of NGC\,1097 (9~pc) is comparable to the sizes of star clusters with masses
$\sim$ 10$^6~M_\odot$ \citep{cal97}; hence, the cluster is quite close to the
nuclear black hole \citep{per93}. The disruption of a
cluster star passing closer to the black hole than its tidal
radius could be the origin of the transient accretion disk, as we have
proposed in \citet{sto93}. Flares in the UV \citep{ren95}
and X-rays \citep{hal04} suggest that we are beginning to witness such events.
Alternatively, mass loss from evolving stars in the cluster
could be the origin of the accretion disk.  In any case,
we may be seeing a rare example of 
the ``starburst-AGN connection'' -- the starburst is responsible
for feeding the AGN! 

Another interesting characteristic of this starburst
is that it is immersed in an obscuring medium.
Its maximum size is also on the order of that predicted for the
torus of the ``Unified Model'' of AGNs \citep{ant93}. Although we cannot
prove that the geometry is that of a circumnuclear torus,
there is no evidence that the nuclear source is obscured,
as we can observe the RIAF + thin-disk continua and the
double-peaked emission-lines with no signs of obscuration;
indeed, the Mg\,II line in the UV is the strongest double-peaked
line of the spectrum. Thus, the starburst {\it may be} within
a dusty toroidal structure which leaves the nuclear source
unobscured. If this is the case, we are observing,
for the first time, a starburst inside a torus, as has
been proposed in the literature as a solution for the
stability problem of the torus: the toroidal structure is supported by
the pressure produced by the evolving starburst inside \citep{wad02,cid95}.

Finally, we point out that the starburst reddening, 
%affecting the starburst,
$A_V = 3$ mag, is not very high. An additional reddening of 2--3
mag would render the starburst undetectable.
Since higher reddening values have been found in the
nuclear regions of many AGNs, our results show that starbursts
may be present, but most will not be detected, at least in the UV.

\acknowledgments

We acknowledge valuable discussions with E. Bica,
C. Bonatto, G. Kriss, C. Leitherer, and P. Hall.
We thank K. Leighly for providing the
Fe\,II spectrum and the referee for valuable suggestions.
This work was supported by NASA grant GO-08684 
from the Space Telescope Science Institute, which is operated by
AURA, Inc., under NASA contract NAS 5-26555,
and the  Brazilian institutions CNPq,
CAPES, and FAPERGS. 
Filippenko is grateful for a Miller Research Professorship at
U.C. Berkeley, and 
Eracleous acknowledges support from NASA LTSA grant 
NAG5-10817.

\clearpage

\end{document}